\documentclass[prl,aps,twocolumn,groupedaddress,floats,final,superscriptaddress,nobibnotes,floatfix]{revtex4-2}

\usepackage{graphicx}
\usepackage{amsmath,amsfonts,amssymb,bm}
\usepackage{braket}
\usepackage{hyperref}
\usepackage{xcolor}

\newcommand{\Tr}{\mathrm{Tr}}
\newcommand{\vk}{\mathbf{k}}

\newcommand{\vA}{\mathbf{A}}
\newcommand{\vR}{\mathbf{R}}

\begin{document}

\title{Superfluid Stiffness and the BKT Transition Temperature in Models for Cuprate Superconductors}

\author{Yueyi Wang}
\thanks{These authors contributed equally to this work.}
\affiliation{Center for Materials Theory, Department of Physics and Astronomy, Rutgers University, Piscataway, New Jersey 08854, USA}
\author{Jiaming Wang}
\thanks{These authors contributed equally to this work.}
\affiliation{Center for Materials Theory, Department of Physics and Astronomy, Rutgers University, Piscataway, New Jersey 08854, USA}
\author{P\"aivi T\"orm\"a}
\affiliation{Department of Applied Physics, Aalto University School of Science, FI-00076 Aalto, Finland}
\affiliation{Department of Physics, Princeton University, Princeton, NJ 08544, USA}
\author{Kristjan Haule}
\email{haule@physics.rutgers.edu}
\affiliation{Center for Materials Theory, Department of Physics and Astronomy, Rutgers University, Piscataway, New Jersey 08854, USA}
\begin{abstract}
The superconducting transition temperature of the cuprates reflects an interplay between pairing strength and phase stiffness. We compute the doping- and temperature-dependent superfluid stiffness of the two-dimensional Hubbard model using cellular dynamical mean-field theory on a $2\times2$ plaquette. We use the impurity natural orbitals renormalization group (iNORG) impurity solver at zero temperature and the bold hybridization quantum Monte Carlo (BHQMC) impurity solver at finite temperature. 
The self-consistent iNORG calculations at zero temperature establish that a gauge-covariant twist of the self-energy, which incorporates vertex corrections beyond the bubble approximation, reproduces the superfluid stiffness to within a few percent.
For cuprate-relevant parameters, the calculated stiffness has the experimentally observed order of magnitude and doping dependence, and increases with hole doping while the superconducting gap decreases. Combining the microscopic stiffness with the BKT universal-jump criterion yields a superconducting dome in which the underdoped regime is limited by low phase stiffness, while the overdoped regime is limited by the weakening of superconductivity. The resulting maximum transition scale occurs near optimal doping and lies in the experimentally relevant range for cuprate superconductors.
\end{abstract}

\maketitle

\textbf{Introduction.} 
Understanding what sets the superconducting transition temperature $T_c$ of the cuprates remains one of the central challenges in quantum materials, both because of their unusually high $T_c$ values and because superconductivity in these materials emerges from strongly correlated electrons near a Mott-insulating state.
Experimental studies over nearly four decades have shown that $T_c$ depends not only on the pairing energy scale but also on the superfluid stiffness of the material~\cite{emerykivelson1995,Uemura1989,UEMURA200323}. In the overdoped regime, the transition temperature of cuprate superconductors decreases as the superconducting gap is reduced, whereas underdoped cuprates exhibit larger superconducting gaps but lower $T_c$ values~\cite{PhysRevLett.80.157,PhysRevB.66.054516,Gomes2007,Boyer2007}. 
In the underdoped regime, experiments have instead found that $T_c$ closely tracks the low-temperature $\mu$SR relaxation rate $\sigma(T\rightarrow 0)$, which is proportional to the inverse square of the penetration depth, $\lambda^{-2}$, and hence to the superfluid stiffness $D_s$~\cite{UEMURA200323,Bernhard,Hg1201andLSCO1,Hg1201andLSCO2,underdopedYBCO}. This trend is consistent with a picture in which pairing persists above \(T_c\), while thermally excited vortices destroy long-range phase coherence at \(T_c\). Nernst measurements on underdoped cuprates support this interpretation~\cite{Ong}. A quantitative microscopic theory of $T_c$ must therefore go beyond determining the pairing amplitude alone and describe both pairing and phase coherence within the same strongly correlated framework~\cite{Kivelson2026}.

Recent progress has made it possible to describe superconductivity in the cuprates within material-specific, ab initio cluster dynamical mean-field theory, yielding superconducting order parameters and gap scales from a microscopic electronic-structure framework~\cite{BenjaminPRX}.
These calculations, however, have so far been restricted to the zero-temperature superconducting state and have not provided a prediction of $T_c$. For conventional phonon-mediated superconductors, controlled first-principles predictions of $T_c$, including a precision many-body treatment of the Coulomb interaction, have recently become possible~\cite{cai2026}. Developing a comparable microscopic approach for the strongly correlated cuprates is therefore an essential next step toward a predictive ab initio theory of cuprate superconductivity.

The empirical separation of pairing and phase-coherence scales suggests that the superconducting transition in cuprates should be viewed not only as the onset of pair formation, but also as the establishment of long-range phase coherence. In the quasi-two-dimensional cuprates, the relevant phase-ordering scale of the CuO$_2$ planes is set by the superfluid stiffness. The Berezinskii--Kosterlitz--Thouless (BKT) criterion~\cite{Kosterlitz,Berezinskii,BKT3} therefore provides a natural way to connect a microscopically calculated superfluid stiffness to the superconducting transition temperature.

In contrast to calculations of the order parameter, microscopic calculations of the superfluid stiffness for cuprates remain comparatively underdeveloped. This is because $D_s$ is not a local measure of pairing but a gauge response to a phase twist~\cite{stiffness_def,ShastrySutherland1990,ScalapinoWhiteZhang}. In a strongly interacting system, this response is renormalized by correlations and requires a consistent treatment of vertex corrections~\cite{Chubukov2024}. 
Previous cluster calculations have addressed the zero-temperature stiffness~\cite{AMTstiffness}, but determining the superconducting transition requires its doping and temperature dependence, with vertex corrections included.

The two-dimensional Hubbard model is the canonical minimal model for doped Mott physics in the CuO$_2$ planes~\cite{ZhangRice1988} and exhibits strong $d$-wave pairing tendencies and superconductivity in the relevant parameter regime~\cite{Maier2005,Gull2013,Jiang2019,shiweiscience,SC_NQS}. Establishing how the superfluid stiffness determines $T_c$ in this model is therefore a crucial first step toward extending material-specific ab initio approaches from the superconducting ground state to a quantitative prediction of the transition temperature. In this work, we compute the doping- and temperature-dependent superfluid stiffness within $2\times2$ cluster dynamical mean-field theory~\cite{DMFT,DMFTRMP}, including vertex corrections to the electromagnetic response. Combined with the BKT criterion, this provides a microscopic framework for determining the phase-ordering temperature of a strongly correlated $d$-wave superconductor.

\textbf{Method.}
As a minimal microscopic model of the CuO$_2$ planes, we study the two-dimensional Hubbard model on the square lattice in the grand-canonical ensemble,
\begin{align}
  \mathcal{K}=-\sum_{  i,j, \sigma} t_{ij}\hat{c}^{\dagger}_{i \sigma} \hat{c}_{j\sigma}+U\sum_i \hat{n}_{i \uparrow}\hat{n}_{i \downarrow}
  -\mu \sum_{i \sigma}\hat{n}_{i\sigma},
  \label{Hubbard}
\end{align}
where $U$ is the local Coulomb repulsion, $\hat{c}_{i\sigma}^\dagger$ and $\hat{c}_{i\sigma}$ are the creation and annihilation operators for an electron at site $i$ with spin $\sigma$, respectively, and $\hat{n}_{i\sigma}=\hat{c}_{i\sigma}^\dagger\hat{c}_{i\sigma}$ is the corresponding number operator. Here $\mu$ is the chemical potential and $t_{ij}$ is the hopping amplitude between sites $i$ and $j$. The bare electron dispersion is modeled using first-, second-, and third-nearest-neighbor hoppings, denoted by $t$, $t'$, and $t''$, respectively. For the cuprate-like model, we use $U/t=8$, $t'/t=-0.3$, and $t''/t=0.2$, representative of cuprate band structures such as YBCO~\cite{Pavarini2001}. For comparison, we also study the nearest-neighbor Hubbard model with $U/t=8$ and $t'=t''=0$. Throughout, we use $t$ as the unit of energy. Physical units are restored when comparing with experiment, as discussed in the Supplemental Material.

The superfluid stiffness is defined from the response of the grand potential to a uniform single-electron $U(1)$ twist~\cite{stiffness_def,ShastrySutherland1990}. On the lattice, the twist is introduced through the Peierls substitution~\cite{Peierls1933} in the hopping term of Eq.~\eqref{Hubbard},
\begin{equation}
t_{ij}
\rightarrow
t_{ij}
e^{-i\mathbf A\cdot(\mathbf r_i-\mathbf r_j)},
\label{phasetwist}
\end{equation}
where $\mathbf r_i$ denotes the position of site $i$. 
The superfluid stiffness per Cu site is then given by the curvature of the grand potential,
\begin{equation}
D_{s,\mu\nu}
=\frac{1}{N_c}
\left.
\frac{d^2 \Omega(\mathbf A)}
{dA_\mu d A_\nu}
\right|_{\mathbf A=0},\label{stiffnessdef}
\end{equation}
where $N_c$ is the number of Cu sites in the cluster. 
The grand potential can be obtained as the stationary value of the Baym--Kadanoff functional~\cite{BKfunctional1,BKfunctional2},
\begin{equation}
\Gamma[G;\mathbf{A}]=\Tr\ln(G)-\Tr\!\left[G_0^{-1}(\vA)G\right]+\Phi[G].\label{BKfunctional}
\end{equation}
For the local Hubbard interaction, the Baym--Kadanoff functional depends explicitly on $\vA$ only through $G_0^{-1}(\vA)$, while the stationary Green's function $G(\vA)$ also depends implicitly on $\vA$. Explicitly, $\Omega(\vA)=\Gamma[G(\vA);\vA]$, where $G(\vA)$ satisfies the stationarity condition
\begin{equation} 
\left.
\frac{\delta\Gamma[G;\mathbf A]}{\delta G}
\right|_{G(\vA)}=0.
\end{equation}
Consequently, the first derivative of the grand potential is given by the partial derivative only:
\begin{align}
&
\frac{d\Omega(\mathbf{A})}{dA_\mu}=
\Tr\!\left[
-\frac{\partial G_0^{-1}(\mathbf{A})}{\partial A_\mu}G(\vA)
\right]\nonumber\\
&=\Tr\!\left[\left(v^0_\mu\otimes\tau_0+\sum_\nu A_\nu T_{\nu\mu}\otimes\tau_3\right)G(\vA)\right],
\label{current}
\end{align}
where 
\begin{eqnarray}
v^0_\mu=\frac{\partial H_\vk}{\partial k_\mu}, \quad T_{\mu\nu}=\frac{\partial^2 H_\vk}{\partial k_\mu\partial k_\nu},
\end{eqnarray}
$\tau_0$ and $\tau_3$ act in Nambu space and the Green's function $G(\vA)$ is computed in the superconducting state in the presence of the finite twist $\vA$.
The physical current density, $J_\mu(\mathbf{A})=-{1}/{N_c}{d\Omega(\mathbf{A})}/{dA_\mu}$, is computed from Eq.~\eqref{current}. 
The superfluid stiffness then follows from Eq.~\eqref{stiffnessdef} as $D_{s,\mu\nu}=-
\left.{d J_\mu(\vA)}/{d A_\nu}
\right|_{\mathbf A=0}$
and is evaluated numerically using a sufficiently small $\vA$. This approach is similar to that of Ref.~\cite{Paivi2017}, where $D_s$ was extracted from DMFT calculations with a finite twist.

 \begin{figure}[b]
  \centering   \includegraphics[width=\columnwidth,trim={7pt 7pt 7pt 7pt},clip]{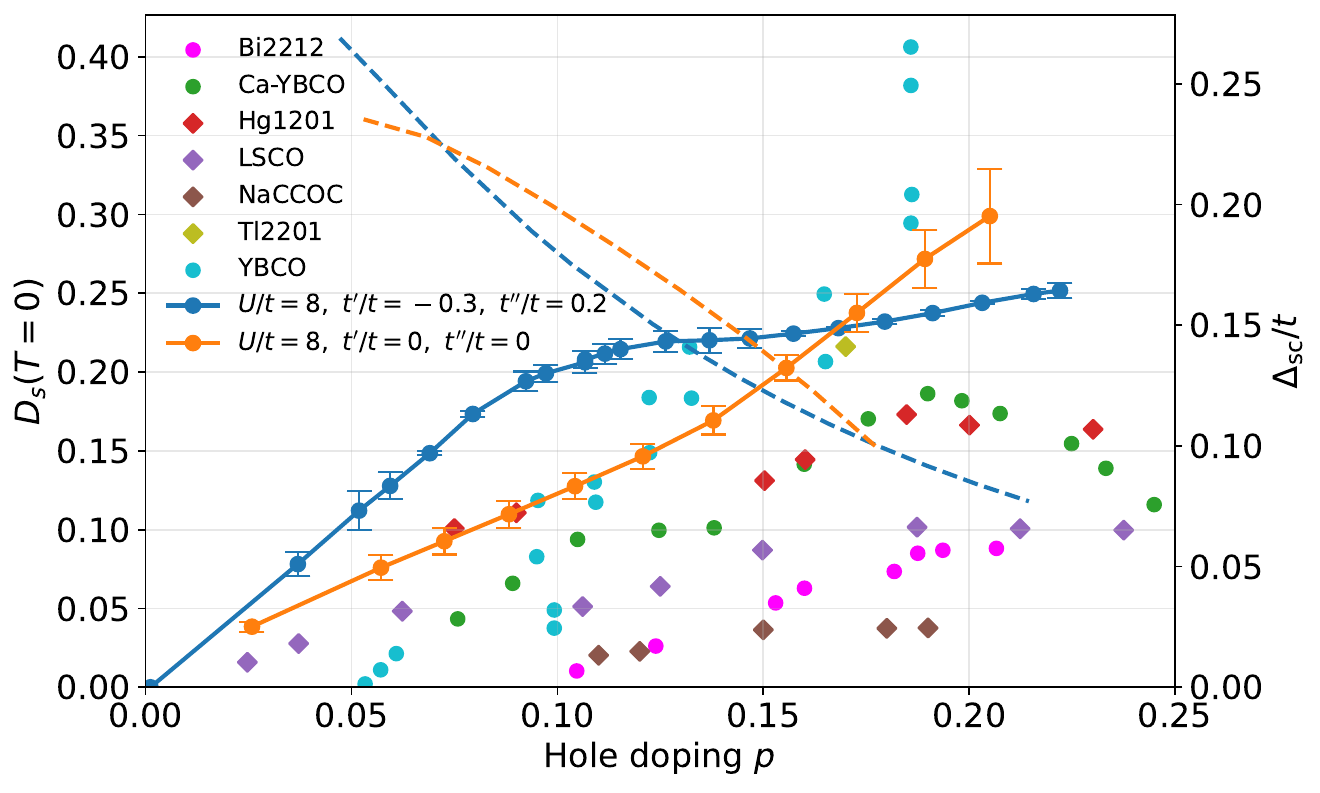}
   \caption{Zero-temperature superfluid stiffness $D_s$ (left axis, solid lines with symbols and error bars) and antinodal superconducting gap $\Delta_{\rm sc}$ (right axis, dashed lines) as functions of hole doping for two sets of hopping parameters at $U/t=8$. The remaining symbols show experimental stiffness values converted to model units~\cite{Bi2212_lambda,Hg1201andLSCO1,Hg1201andLSCO2,CCOC_lambda,Tl2201_lambda,underdopedYBCO,YBCO2,YBCO3,YBCO4}.}
   \label{T0stiffness}
 \end{figure}

In this study, the interacting Green's function is computed within cellular dynamical mean-field theory (cDMFT)~\cite{Gabi_CDMFT} on a $2\times2$ cluster, allowing for a $d$-wave superconducting solution.
\begin{figure*}[htbp]
  \centering
  \begin{minipage}{0.32\textwidth}
    \centering
    \llap{\raisebox{-1.2em}{\textbf{(a)}\hspace{0.1em}}}%
    \includegraphics[width=\linewidth,trim={5pt 10pt 5pt 5pt},clip]{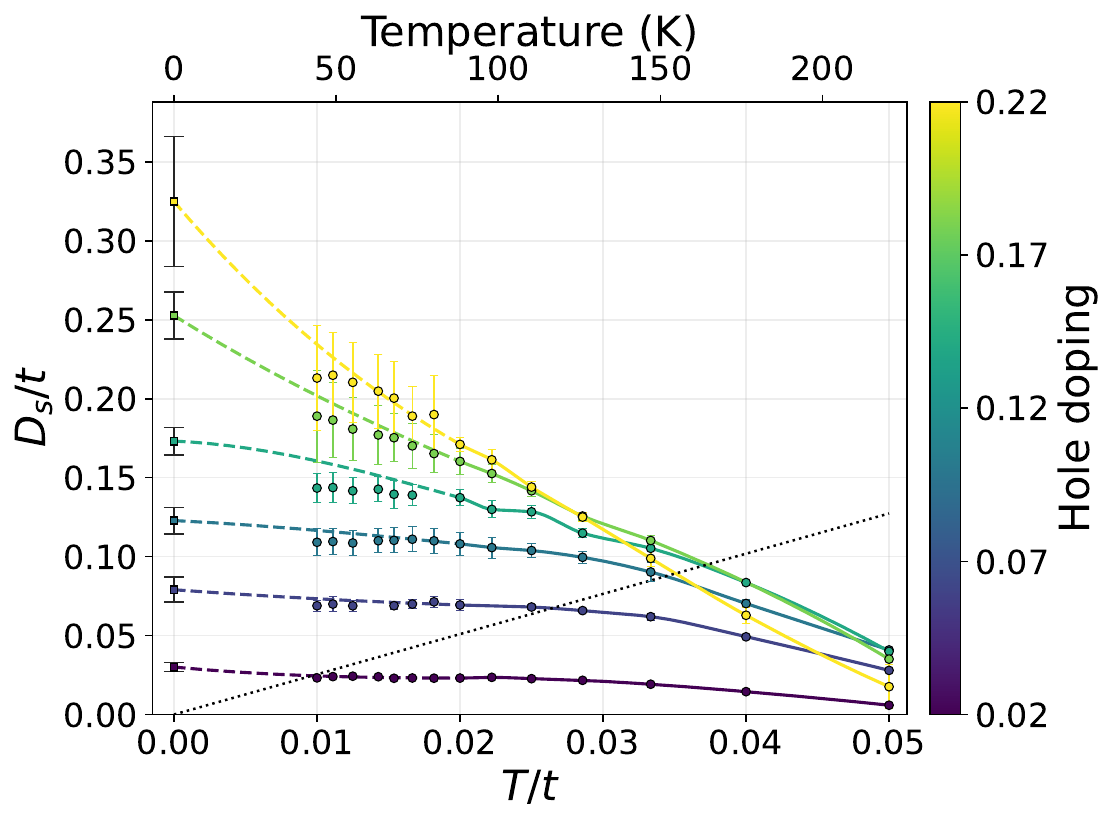}
  \end{minipage}
  \begin{minipage}{0.32\textwidth}
    \centering
    \llap{\raisebox{-1.2em}{\textbf{(b)}\hspace{0.1em}}}%
    \includegraphics[width=\linewidth,trim={5pt 10pt 5pt 5pt},clip]{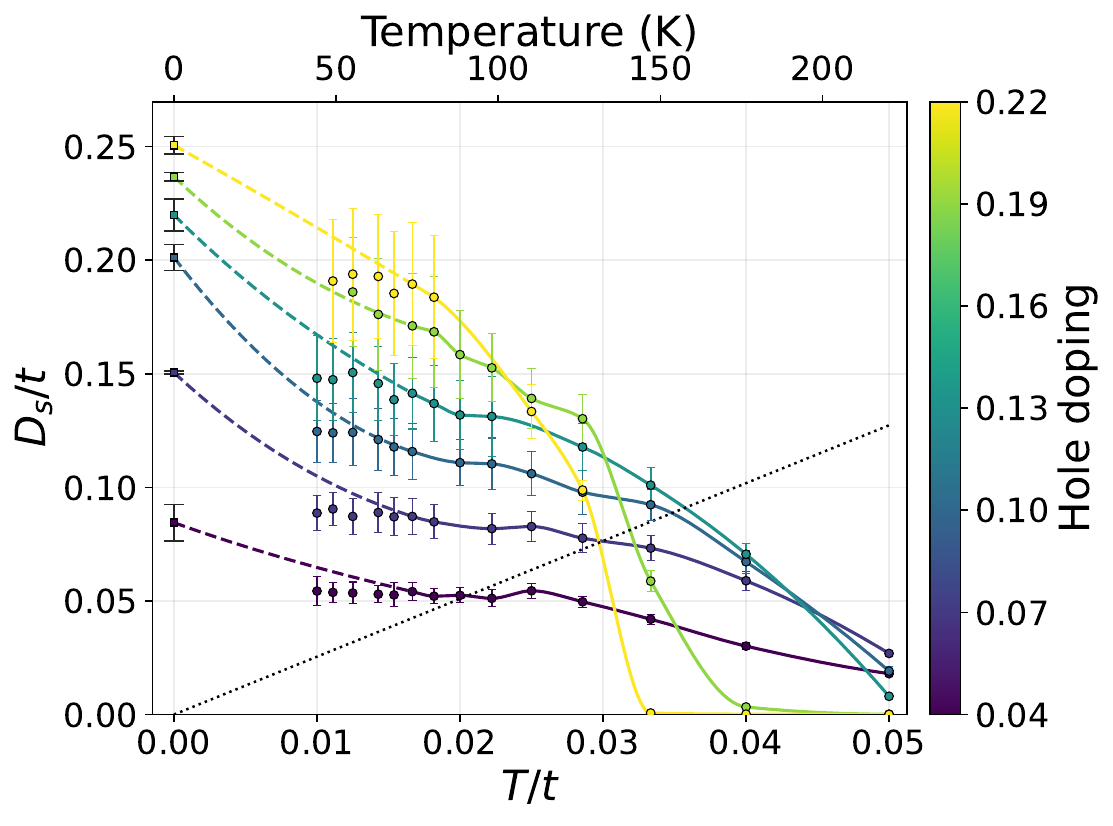}
  \end{minipage}
  \begin{minipage}{0.32\textwidth}
    \centering
    \llap{\raisebox{-1.2em}{\textbf{(c)}\hspace{0.1em}}}%
    \includegraphics[width=\linewidth,trim={5pt 10pt 5pt 5pt},clip]{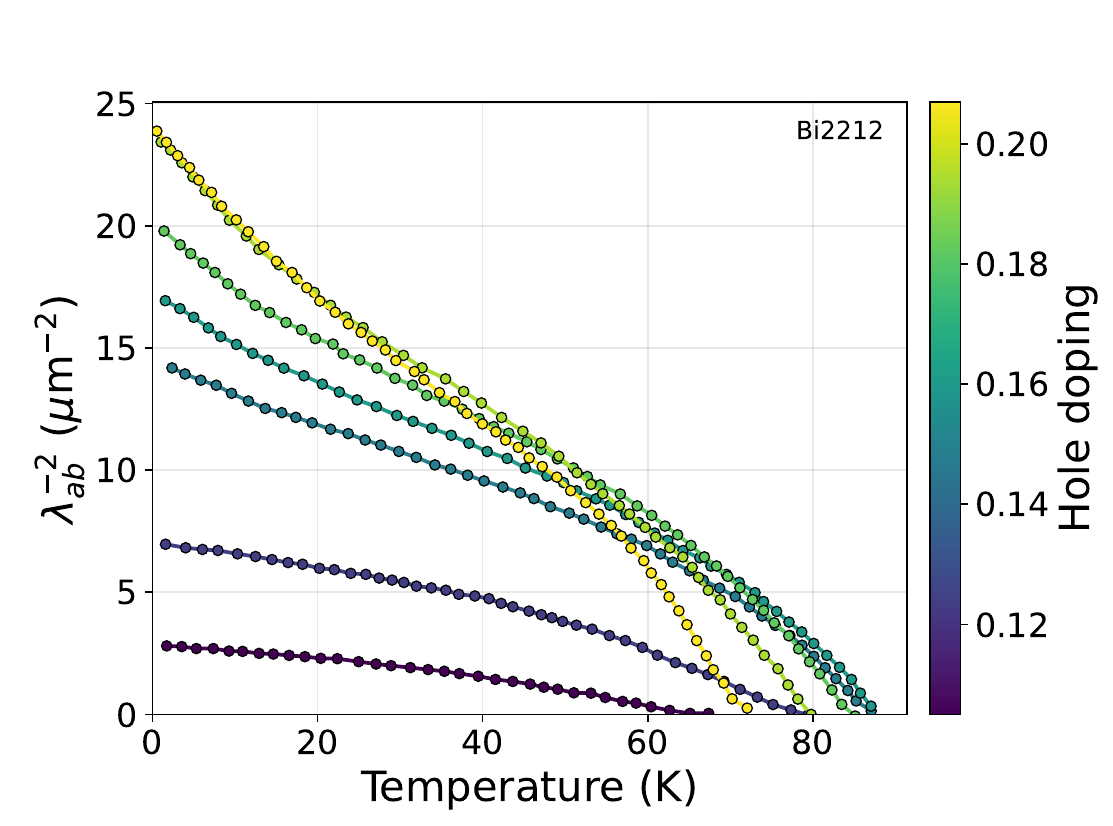}
  \end{minipage}
  \caption{Finite-temperature superfluid stiffness of the Hubbard model with (a) $U/t=8$, $t'/t=t''/t=0$ and (b) $U/t=8$, $t'/t=-0.3$, $t''/t=0.2$. (c) Experimentally measured inverse squared penetration depth $\lambda^{-2}\propto D_s$ in Bi2212~\cite{Bi2212_lambda}.  Dashed lines connect the $T=0$ iNORG data points to the finite-temperature data points as guides to the eye. Black dotted lines indicate the relation $y=\frac{8}{\pi}x$, which corresponds to the BKT criterion. Temperatures in Kelvin are estimated using $t=0.38\,\mathrm{eV}$.}
  \label{DsT}
\end{figure*}
We solve the plaquette cDMFT equations with the impurity natural orbitals renormalization group (iNORG) impurity solver~\cite{iNORG,NORG} at zero temperature and with the fourth-order bold hybridization quantum Monte Carlo (BHQMC) impurity solver~\cite{BHQMC} at finite temperature. 
Evaluating the current $\mathbf J(\mathbf A)$ in Eq.~\ref{current} requires the self-energy in the presence of the phase twist. We construct it by applying the local gauge transformation associated with the Peierls substitution to the converged zero-field self-energy. For the full Nambu matrix, this gives
\begin{eqnarray}
\Sigma_{ij}(\vA)\approx
V_i
\begin{pmatrix}
\Sigma^{11}_{ij}(\vA=0)  & \Sigma^{12}_{ij}(\vA=0)\\
\Sigma^{21}_{ij}(\vA=0)  & \Sigma^{22}_{ij}(\vA=0)
\end{pmatrix}
V_j^\dagger,
\label{twistingapprox}
\end{eqnarray}
where $V_i(\vA)=e^{-i\tau_3\vA\cdot\boldsymbol{\rho}_i}$,

and $\boldsymbol\rho_i$ is the position of site $i$ within the unit cell.
For a local pairing field, 
Eq.~\ref{twistingapprox} reduces to the intracell phase factor 
$\Delta_j\to\Delta_j e^{-2i\vA\cdot\boldsymbol{\rho}_j}$ used in the multiband BCS treatment~\cite{Lieb_stiffness}. 
The twist generates a nonzero $d\Sigma/d\mathbf A$, incorporating vertex corrections beyond the dressed bubble from the zero-field solution.
We benchmark Eq.~\ref{twistingapprox} at zero temperature by solving the Peierls-substituted plaquette cDMFT equations self-consistently with iNORG at each $\mathbf A$ and numerically differentiating the resulting current. For both parameter sets, the gauge-covariant twist reproduces the fully self-consistent stiffness to within about $10\%$ across the doping range tested (Fig.~4). This agreement extends to the underdoped regime, where the dressed bubble substantially overestimates the stiffness. The construction requires self-consistency only at zero field, making it efficient to use for the stiffness calculations at both zero and finite temperature.

Having obtained the microscopic cDMFT stiffness $D_s(T)$, we use it as the phase rigidity governing long-wavelength fluctuations of the superconducting phase. In two dimensions, the loss of phase coherence occurs through the unbinding of vortex-antivortex pairs, with the transition determined by the BKT universal-jump criterion~\cite{Kosterlitz,Berezinskii,BKT3},
\begin{equation}
T_c^{\rm BKT}=\frac{\pi}{8}D_{s,xx}(T_c^{\rm BKT}).
\label{BKT}
\end{equation}

\textbf{Results and Discussion.}
To establish how the superfluid stiffness and superconducting gap evolve with hole doping $p$, we calculate the zero-temperature stiffness $D_s(T=0)$ and estimate the antinodal quasiparticle gap scale as
$\Delta_{\rm sc}=Z_{(0,\pi)}|\Sigma^{\rm an}_{(0,\pi)}(i\omega_n\to0)|$,
where $\Sigma^{\rm an}_{(0,\pi)}$ is the anomalous self-energy at the antinodal momentum and $Z_{(0,\pi)}$ is the corresponding quasiparticle weight, determined from the low-frequency normal self-energy. 
Both quantities are obtained using the iNORG impurity solver (see Supplemental Material for details) and are shown in Fig.~\ref{T0stiffness}, together with experimental stiffness values inferred from measurements of the penetration depth $\lambda$ and the $\mu$SR relaxation rate $\sigma$. 
Both parameter sets reproduce the experimentally observed increase in $D_s(T=0)$ and decrease in the antinodal gap with increasing hole doping~\cite{Bi2212_lambda,Hg1201andLSCO1,doi:10.1126/science.1174338}. Including the cuprate-relevant $t'$ and $t''$ considerably enhances the stiffness in the underdoped and intermediate-doping regimes. The sensitivity of the stiffness scale to the hopping parameters is consistent with the variation among cuprate families seen in Fig.~\ref{T0stiffness}. As shown in Supplemental Material, across both parameter sets, $D_s$ correlates with the product of the bare-band Drude weight and the antinodal quasiparticle weight.

The calculated stiffness has the correct order of magnitude and reproduces the experimentally observed doping dependence, although its magnitude is generally larger than the measured values. 
The difference may reflect disorder in the experiment, electronic correlations beyond next-nearest neighbors neglected here, and material-specific effects of crystal structure and chemistry that can be addressed within an ab initio framework~\cite{BenjaminPRX}.

 \begin{figure}[bp]
  \centering
   \includegraphics[width=\columnwidth,trim={0pt 0pt 0pt 0pt},clip]{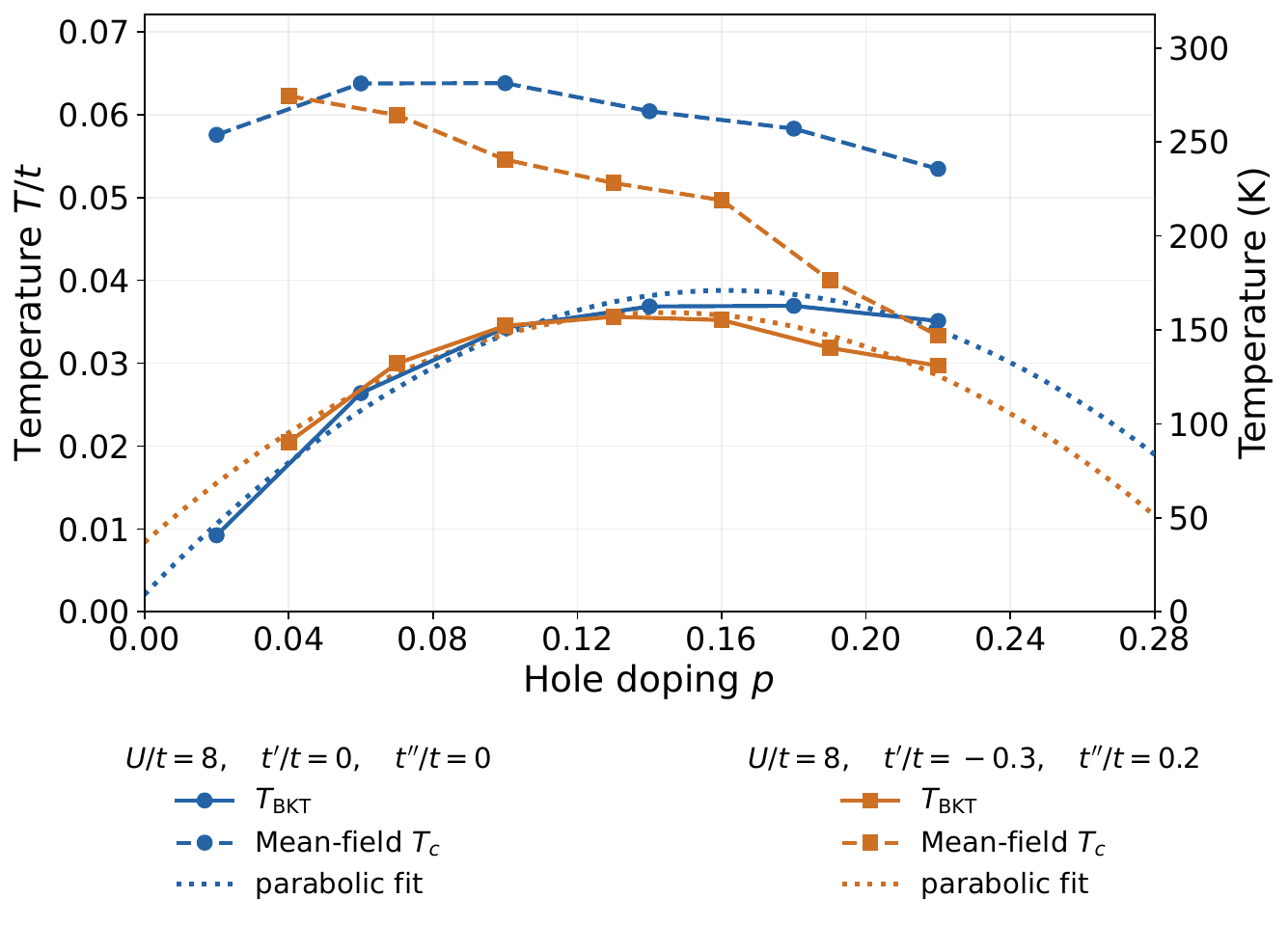}
   \caption{Superconducting transition temperature of the Hubbard model with different parameter sets. The solid line shows the BKT transition temperature, whereas the dashed line shows the $T_c$ defined by the vanishing of the anomalous self-energy in the cDMFT solution, i.e., the mean-field $T_c$. Dotted lines are parabolic fits to the BKT transition temperature.}
   \label{Tc}
 \end{figure}

Having established the zero-temperature stiffness scale, we now turn to its temperature dependence, which determines the BKT phase-ordering scale. Figure~\ref{DsT} shows the finite-temperature results obtained with the fourth-order BHQMC impurity solver, together with the $T=0$ iNORG results discussed above. To illustrate the low-temperature trends, the two datasets are connected by dashed lines as guides to the eye. Since the zero- and finite-temperature calculations employ different impurity solvers, some quantitative difference between the $T=0$ iNORG results and the BHQMC results is expected. Nevertheless, the two solvers agree qualitatively for both parameter sets.

In both Hubbard models, the superfluid stiffness increases monotonically with hole doping at low temperature, e.g., $T=0.01$, consistent with the $T=0$ results. Its temperature dependence, however, changes markedly across the superconducting dome. In the underdoped regime, $D_s(T)$ decreases only weakly at low temperature, followed by a more rapid suppression as the superconducting transition is approached. In contrast, the overdoped regime exhibits a substantially stronger temperature dependence already at low temperature. For example, in the nearest-neighbor Hubbard model at $22\%$ hole doping, $D_s(T)$ decreases approximately linearly from $T=0.01$ until superconductivity disappears near $T\simeq0.05$. Such a linear-in-$T$ suppression, which originates from thermally excited quasiparticles near the gap nodes, is characteristic of a $d$-wave BCS superconductor, with a slope governed by the ratio $v_F/v_\Delta$ of the nodal Fermi and gap velocities. A similar behavior is found in the overdoped cuprate-like model, followed by a sharper suppression near the transition. These characteristic temperature dependences are qualitatively consistent with experimental observations in LSCO~\cite{Hg1201andLSCO2} and Bi2212~\cite{Bi2212_lambda}.

Using the microscopic superfluid stiffness computed within plaquette cDMFT together with the BKT universal-jump criterion, we obtain the phase-ordering temperatures shown in Fig.~\ref{Tc}, along with the cDMFT transition temperature at which the superconducting solution disappears. 
In the nearest-neighbor model, the cDMFT transition temperature peaks in the underdoped regime at a value close to the $0.06\,t$ reported in previous cluster-DMFT studies~\cite{AMTorderparams}. In the cuprate-like model, it continues to rise with underdoping and reaches a comparable temperature scale near $p\simeq 0.04$, where the superconducting solution terminates.

Phase fluctuations, as encoded in the BKT criterion, suppress the transition temperature much more strongly in the underdoped regime, where the phase stiffness is small, than on the overdoped side. Consequently, the maximum of the BKT transition scale shifts toward higher doping, occurring near $17\%$ hole doping for the nearest-neighbor model and $13\%$ for the cuprate-like model, with corresponding maximum temperatures of approximately $160\,\mathrm{K}$ and $150\,\mathrm{K}$, respectively. 
These values are somewhat higher than the $\sim90\,\mathrm{K}$ transition temperatures characteristic of many single-layer optimally doped cuprates, but nevertheless place the phase-ordering scale of the microscopic Hubbard model in the experimentally relevant temperature range.

Experimentally, cuprates with smaller $|t'|/t$ and $t''/t$ generally exhibit lower $T_c$ values~\cite{Pavarini2001}. The weaker superconducting tendency of the nearest-neighbor Hubbard model~\cite{PRXnoSCtonly} has been associated with the formation of filled stripes~\cite{shiweiscience} and with competition between superconducting and antiferromagnetic states~\cite{SCAF2008}. These competing states are not captured by our $2\times2$ plaquette cDMFT calculation, which is restricted to the uniform $d$-wave superconducting solution and cannot represent the long-period stripe states found in larger-scale studies. 
The comparatively high $T_c^{\rm BKT}$ obtained for the nearest-neighbor model should therefore be interpreted as the phase-ordering scale within this restricted uniform superconducting solution and may overestimate the superconducting tendency of the unrestricted model.

\textbf{Conclusion.}
We computed the doping- and temperature-dependent superfluid stiffness of the two-dimensional Hubbard model using cDMFT on a $2\times2$ plaquette.
At zero temperature, we obtained the full vertex-corrected stiffness within plaquette cDMFT by solving the problem self-consistently in the presence of the Peierls field with iNORG. A gauge-covariant twist of the zero-field self-energy reproduces this stiffness to within about $10\%$ across the doping range and provides an efficient way to include vertex corrections beyond the dressed bubble in the finite-temperature BHQMC calculations. 
For cuprate-relevant parameters, the microscopic stiffness has the correct order of magnitude and broadly reproduces the experimentally observed doping and temperature dependence, while the antinodal gap evolves oppositely, decreasing with increasing hole doping. Combining the calculated stiffness with the BKT universal-jump criterion produces a superconducting dome: the transition scale is limited by the small phase stiffness on the underdoped side and by the weakening of superconductivity on the overdoped side.
Our BKT estimate uses the microscopic cDMFT stiffness directly and does not include its further renormalization by long-wavelength vortex-antivortex fluctuations, which can reduce the actual phase-ordering temperature. Competing orders, disorder, and interlayer coupling are likewise beyond the present two-dimensional single-band description. Nevertheless, our results demonstrate that the strongly correlated Hubbard model captures the complementary evolution of pairing and phase stiffness underlying the cuprate superconducting dome.
More broadly, the present approach provides a route for extending material-specific ab initio cluster-DMFT calculations beyond the superconducting ground state toward a quantitative determination of the superconducting transition temperature.

\textbf{Acknowledgments.}
This work was supported by NSF DMR-2233892, by a grant from the Simons Foundation (SFI-MPS-NFS-00006741-06, K.H.; SFI-MPS-NFS-00006741-12, P.T.) in the Simons Collaboration on New Frontiers in Superconductivity, and by a collaboration between The Kavli Foundation, Klaus Tschira Stiftung, and Kevin Wells, and by the Jane and Aatos Erkko Foundation, the Keele Foundation, and the Magnus Ehrnrooth Foundation, as part of the SuperC collaboration. This work was supported by the Research Council of Finland under project number 374169 and is part of the Finnish Centre of Excellence in Quantum Materials (QMAT). The authors acknowledge the Frontera supercomputer at the Texas Advanced Computing Center (TACC) at The University of Texas at Austin, which is supported by National Science Foundation Grant No. OAC-1818253.

\bibliography{SFstiffness}

\clearpage
\onecolumngrid
\section{End Matter}
\twocolumngrid
\subsection{The Linear Response Theory of Superfluid Stiffness}
In this section, we connect our formulation of the superfluid stiffness to the conventional linear-response treatment and show that Eq.~\eqref{twistingapprox} provides a gauge-covariant approximation to the self-energy response $d\Sigma/d\vA$. 
In the context of cellular DMFT and superfluid stiffness, two Fourier-transform conventions are in common use,
\begin{align}
c_{\vk i}&=\sum_{\vR}e^{-i\vk\cdot\vR}\,c_{\vR i},
\label{Fourier1}\\
c_{\vk i}&=\sum_{\vR}e^{-i\vk\cdot(\vR+\boldsymbol{\rho}_i)}\,c_{\vR i},
\label{Fourier}
\end{align}
where $\boldsymbol{\rho}_i$ and $\vR$ labels the position of site $i$ within the unit cell and to the unit cell, respectively. 
In standard cDMFT calculations the former convention is preferred: since the self-energy has no intercell components, its Fourier transform carries no explicit phase factor. However, the latter convention is used in the following discussion, since it allows the Peierls substitution to be represented as a shift in momentum space:
\begin{equation}
H(\vk,\vA)=H(\vk+\vA,0),
\end{equation}
where $H(\vk,\vA)$ denotes the Peierls-substituted tight-binding Hamiltonian in the presence of the phase twist $\vA$ introduced in Eq.~\eqref{phasetwist}. In the following, we use the shorthand $H_{\vk+\vA}\equiv H(\vk+\vA,0)$.

The noninteracting Green's function in Nambu space is
\begin{eqnarray}
G_{0}^{-1}=\begin{pmatrix}
i\omega_n+\mu-H_{\vk+\vA} & 0\\
0 & i\omega_n-\mu+H^*_{-\vk+\vA}
\end{pmatrix}.
\end{eqnarray} 
Its derivative with respect to the phase twist is
\begin{eqnarray}
&-\frac{\partial G_0^{-1}}{\partial A_\nu}=
\begin{pmatrix}
\frac{\partial H_{\vk+\vA}}{\partial A_\nu} & 0\\
0 & -\frac{\partial{H^*_{-\vk+\vA}}}{\partial A_\nu}
\end{pmatrix}.
\end{eqnarray}
Time-reversal symmetry, 
$H^*_{-\vk}=H_{\vk}$, together with an expansion to first order in $\vA$, gives
\begin{eqnarray}
-\frac{\partial G_0^{-1}}{\partial A_\nu}= v^0_\nu(\vk)\otimes\tau_0 + [\sum_{\mu}A_{\mu}  T_{\mu\nu}(\vk) ]\otimes\tau_3,
\end{eqnarray}
where
\begin{eqnarray}
v^0_\mu(\vk)=\frac{\partial H_\vk}{\partial k_\mu}, \quad T_{\mu\nu}(\vk)=\frac{\partial^2 H_\vk}{\partial k_\mu\partial k_\nu}.
\end{eqnarray}
Using these results, Eq.~\eqref{current} can be rewritten as
\begin{align}
J_\mu(\vA)=-\frac{1}{N_c}\Tr([\sum_{\nu} (A_{\nu} T_{\nu\mu})\otimes\tau_3+v^0_\mu\otimes\tau_0]G(\vA)),\label{currentendmatter}
\end{align}
and therefore
\begin{align}
&D_{s,\mu\nu}=-\left.\frac{d J_\mu}{dA_\nu}\right|_{\vA=0} \nonumber \\
&= \frac{1}{N_c}\left.\Tr\left(T_{\nu\mu}\otimes\tau_3\, G(\vA)+v^0_\mu\otimes\tau_0\frac{dG(\vA)}{dA_\nu}\right)\right|_{\vA=0}.
\label{stiffness}
\end{align}
The first and second terms are then the diamagnetic and paramagnetic contributions, respectively.

The identity $G G^{-1}=1$ 
allows the derivative of the Green's function in the paramagnetic term to be written as
\begin{align}
\left.\frac{dG}{dA_\nu}\right|_{\vA=0}
&=-\left.G\,\frac{dG^{-1}}{dA_\nu}\,G\right|_{\vA=0}\nonumber\\
&=\left.G\left(-\frac{dG_0^{-1}}{dA_\nu}+\frac{\delta\Sigma}{\delta G}\frac{dG}{dA_\nu}\right)G\right|_{\vA=0}\nonumber\\
&=\left.G\left(v^0_\nu\otimes\tau_0+\frac{\delta\Sigma}{\delta G}\frac{dG}{dA_\nu}\right)G\right|_{\vA=0}.
\end{align}

The first term gives the so-called ``bubble approximation'' to the
paramagnetic contribution,
\begin{eqnarray}
D^{\rm bubble}_{\mathrm{para},\mu\nu}
=
\frac{1}{N_c}
\Tr\left[
(v^0_\mu\otimes\tau_0)
G
(v^0_\nu\otimes\tau_0)
G
\right].
\end{eqnarray}
The second term generates the vertex correction, which can be expressed
through the Bethe--Salpeter equation
\begin{eqnarray}
\Lambda_\nu
=
v^0_\nu\otimes\tau_0
+
\frac{\delta\Sigma}{\delta G}
G\Lambda_\nu G.
\end{eqnarray}
Here,
\begin{eqnarray}
\Lambda_\nu=-\frac{dG^{-1}}{dA_\nu}
\end{eqnarray}
is the dressed three-leg current vertex, while
$\delta\Sigma/\delta G$ is the particle-hole irreducible four-point
vertex. The full paramagnetic contribution can therefore be written as
\begin{eqnarray}
D_{\mathrm{para},\mu\nu}
=
\frac{1}{N_c}
\Tr\left[
(v^0_\mu\otimes\tau_0)
G\Lambda_\nu G
\right],
\end{eqnarray}
which reduces to the bubble approximation when
$\Lambda_\nu=v^0_\nu\otimes\tau_0$. In multiside systems, terms going beyond the bubble approximation  are important even for the mean-field superfluid stiffness~\cite{Huhtinen2022}.

Direct evaluation of the vertex correction through the Bethe--Salpeter equation requires the particle-hole irreducible four-point vertex $\delta\Sigma/\delta G$, whose computation is considerably more demanding than solving the interacting problem in the presence of a small finite phase twist $\vA$. We therefore obtain the superfluid stiffness by evaluating
\begin{eqnarray}
G(\vA)=\left[G_0^{-1}(\vA)-\Sigma(\vA)\right]^{-1}
\end{eqnarray}
at a small enough $\vA$ and numerically differentiating the current in Eq.~\eqref{currentendmatter}. When $\Sigma(\vA)$ is obtained from a fully self-consistent solution at a small enough $\vA$, this procedure includes the complete self-energy response within plaquette cDMFT, and hence the corresponding vertex corrections, without explicitly computing the four-point vertex. To assess the importance of the self-energy response and the accuracy of the twisting approximation, we compare the following three calculations in Fig.~\ref{twistbenchmark}:

\begin{figure}[t]
    \centering

    \llap{\raisebox{-1.2em}{\textbf{(a)}\hspace{0.5em}}}%
    \includegraphics[width=\linewidth,trim={22pt 10pt 10pt 15pt},clip]{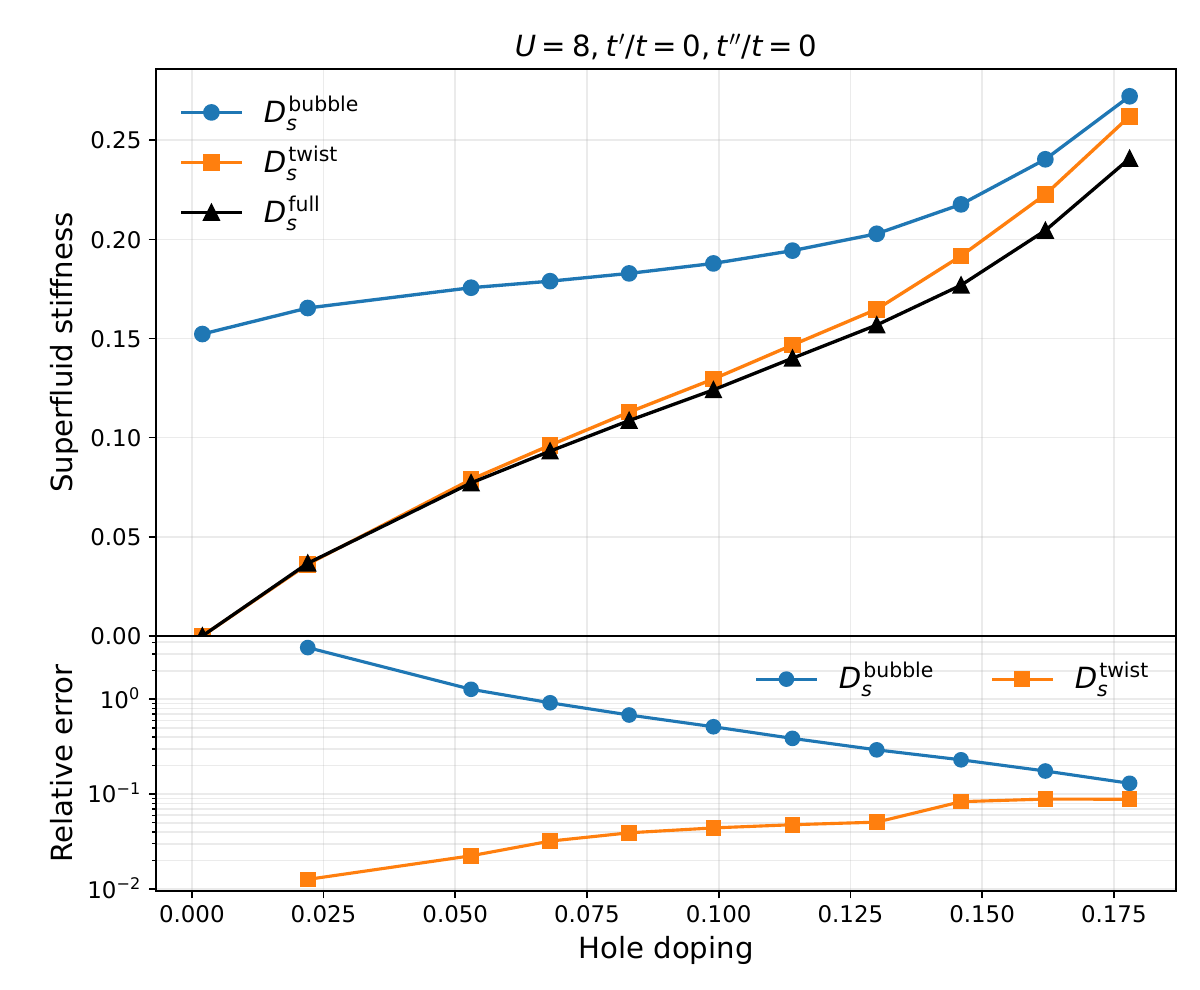}

    \vspace{0.1em}

    \llap{\raisebox{-1.2em}{\textbf{(b)}\hspace{0.5em}}}%
    \includegraphics[width=\linewidth,trim={22pt 10pt 10pt 15pt},clip]{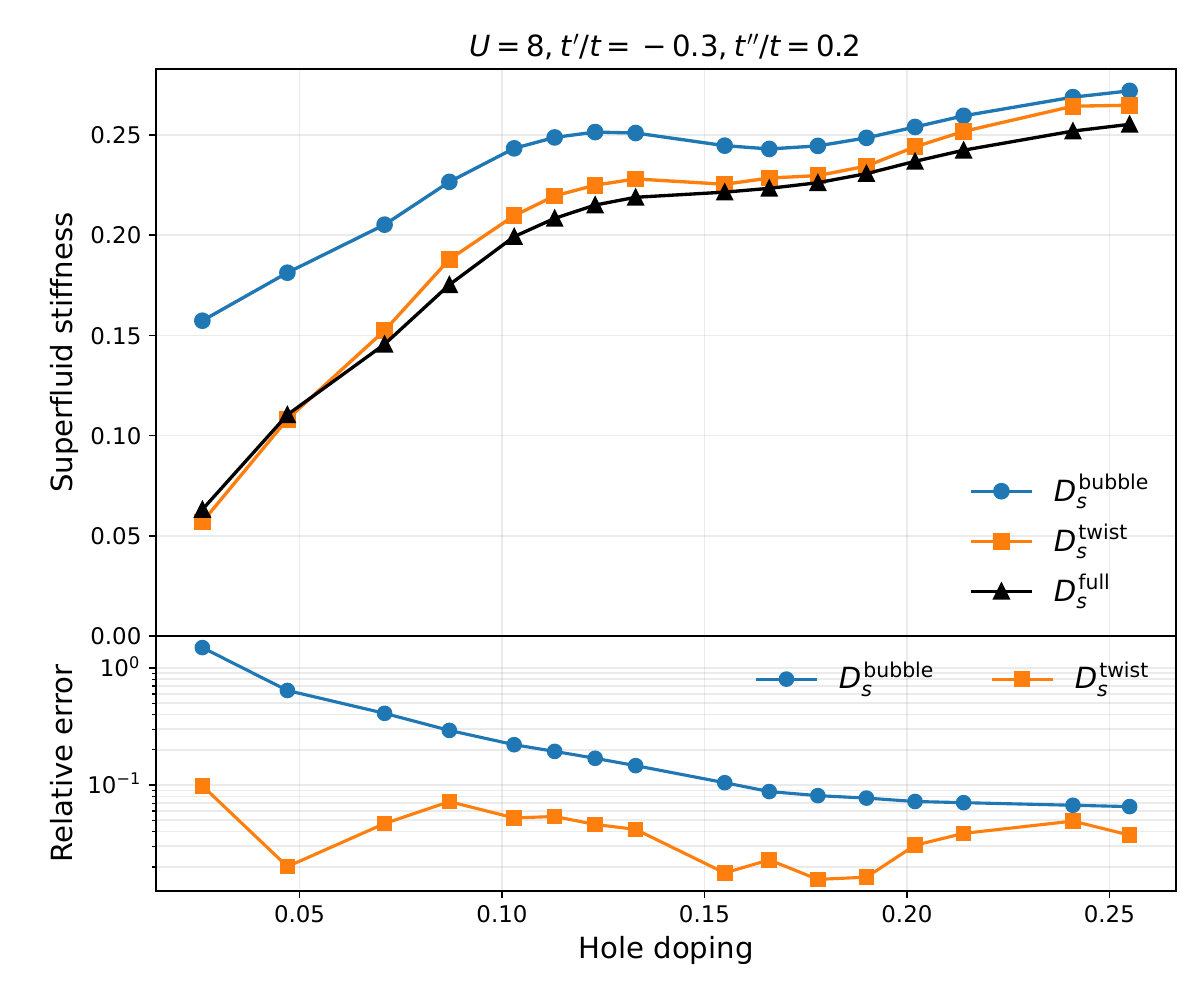}

    \caption{
    Zero-temperature superfluid stiffness obtained in the bubble and twisting approximations, compared with the fully self-consistent cDMFT result in the presence of the Peierls substitution, for the Hubbard model with parameters
    (a) $U/t=8$, $t'/t=t''/t=0$ and
    (b) $U/t=8$, $t'/t=-0.3$, $t''/t=0.2$. The lower panels show the relative errors with respect to the fully self-consistent calculations.
    }
    \label{twistbenchmark}
\end{figure}
\begin{itemize}
\item \emph{Bubble approximation.} The self-energy is held fixed at its zero-field value, $\Sigma(\vA)\approx\Sigma(0)$, so that $d\Sigma/d\vA=0$. The paramagnetic response is therefore evaluated at the dressed-bubble level, yielding $D_s^{\rm bubble}$.
\item \emph{Twisting approximation.}
The self-energy response is approximated by the gauge-covariant transformation
\begin{eqnarray}
\Sigma_{ij}(\vA)= V_i(\vA)\Sigma_{ij}(0)V_j^\dagger(\vA),
\end{eqnarray}
as defined in Eq.~\eqref{twistingapprox}. This retains a nonzero $d\Sigma/d\vA$ and yields $D_s^{\rm twist}$ without requiring a new self-consistent impurity solution at each $\vA$.
\item \emph{Self-consistent response.}
The impurity problem and the cDMFT self-consistency condition are solved anew at each $\vA$, yielding the fully self-consistent $\Sigma(\vA)$ and the corresponding stiffness $D_s^{\rm full}$. This calculation includes the complete self-energy response to the phase twist within plaquette cDMFT. We perform this calculation at zero temperature using the iNORG impurity solver.
\end{itemize}

Fig.~\ref{twistbenchmark} benchmarks the bubble and twisting approximations against the fully self-consistent response at zero temperature. All three calculations in this benchmark use a single bath-fitting setup, whereas the values and error bars in Fig.~\ref{T0stiffness} combine several (see Supplemental Material), so the numbers differ slightly from Fig.~\ref{T0stiffness}.
For both sets of Hubbard-model parameters, the bubble approximation substantially overestimates the superfluid stiffness toward half filling and, in particular, fails to recover the vanishing stiffness as the hole doping $p\rightarrow0$. In contrast, $D_s^{\rm twist}$ closely follows the fully self-consistent $D_s^{\rm full}$ throughout the superconducting doping range. The gauge-transformation law underlying the twisting construction is independent of temperature and of the impurity solver. Together with the zero-temperature benchmark, this indicates that the gauge-covariant twisting approximation captures the dominant self-energy response within plaquette cDMFT. We therefore use the twisting approximation for finite temperature results in the main text.

\clearpage
\onecolumngrid

\section{Supplemental Material}
\renewcommand{\thefigure}{S\arabic{figure}}
\renewcommand{\theHfigure}{S\arabic{figure}}
\setcounter{figure}{0}
\renewcommand{\theequation}{S\arabic{equation}}
\renewcommand{\theHequation}{S\arabic{equation}}
\setcounter{equation}{0}
\renewcommand{\thetable}{S\arabic{table}}
\renewcommand{\theHtable}{S\arabic{table}}
\setcounter{table}{0}
\subsection{Units of Superfluid Stiffness}
To compare our theoretical results with experiment, we convert the dimensionless two-dimensional stiffness per Cu site into the corresponding three-dimensional electromagnetic response, which can then be related to the London penetration depth or the $\mu$SR relaxation rate.

Since the phase accumulated under a boundary twist is dimensionless, the single-electron phase gradient $\vA_{\rm th}$ used in the model and the physical vector potential $\vA_{\rm ex}$ (in SI units) are related by
\begin{equation}
\vA_{\rm th}=\vA_{\rm ex}\frac{ae}{\hbar},
\end{equation}
where $a$ is the distance between neighboring Cu atoms in the CuO$_2$ plane, $e$ is the magnitude of the electron charge and $\hbar$ is the reduced Planck constant.

Another difference between the theoretical and experimental units lies in the normalization of the grand potential. In Eq.~\eqref{stiffnessdef}, the theoretical superfluid stiffness is defined from the grand potential per Cu site in the CuO$_2$ plane. In contrast, the experimental electromagnetic response is defined from the grand-potential density per unit volume, denoted by $\Omega_{\rm ex}$. Therefore,
\begin{align}
	D_{s,{\rm ex}}=\frac{d^2\Omega_{\rm ex}}{dA_{\rm ex}^2}=\frac{\frac{t}{a^2 d}}{(\frac{\hbar}{ea})^2}\frac{d^2\Omega_{\rm th}}{dA_{\rm th}^2}=\frac{te^2}{\hbar^2d}D_{s,{\rm th}}.
	\label{conversion}
\end{align}

Here, $t$ is the nearest-neighbor hopping amplitude, which sets the energy scale of the model and must be expressed in SI energy units (joules) in Eq.~\eqref{conversion}, and $d$ is the $c$-axis lattice constant divided by the number of CuO$_2$ planes in the unit cell. This choice converts the bulk response to a mean sheet stiffness per plane. The in-plane lattice constant $a$ cancels in Eq.~\eqref{conversion}.
If $n$ planes share a common phase and form an effectively isolated two-dimensional block, their total sheet stiffness is $nD_{s,xx}$, where $D_{s,xx}$ is the mean stiffness per plane. In this limit the BKT criterion reads
\begin{equation}
T_c^{\rm BKT}=\frac{n\pi}{8}D_{s,xx}(T_c^{\rm BKT}).
\label{eq:BKTblock}
\end{equation}
The stiffness is evaluated just below the transition. For example,
$n=2$ if the two planes in a Bi2212 bilayer form such a block.

In the transverse London gauge, the electromagnetic current response is $j_\mu=-D_{s,\mu\nu}A_\nu$. Combining this relation with $\nabla\times\mathbf B=\mu_0\mathbf j$ and $\mathbf B=\nabla\times\mathbf A$, one obtains $\lambda^{-2}=\mu_0D_s$. The relation between the penetration depth and the theoretical stiffness is therefore
\begin{align}
\lambda^{-2}
=\frac{\mu_0te^2}{\hbar^2d}D_{s,{\rm th}}.
\label{units}
\end{align}

Taking the representative values $d\sim0.75\,\mathrm{nm}$ and $t\sim0.4\,\mathrm{eV}$ gives $\mu_0te^2/(\hbar^2d)\sim250\,\mu\mathrm{m}^{-2}$. Consequently, a theoretical stiffness of $D_{s,{\rm th}}=0.2$ corresponds to a physical value of approximately $\lambda^{-2}=50\,\mu\mathrm{m}^{-2}$. The material-specific values of $d$ and $t$ are listed in Table~\ref{tab:parameters}.

\begin{table}[h]
\centering
\caption{Material-specific parameters used to convert experimental data to model units: nearest-neighbor hopping $|t|$, interplane spacing $d$, and the resulting conversion factor $\mu_0te^2/\hbar^2d$ of Eq.~\eqref{units}.}
\label{tab:parameters}
\begin{ruledtabular}
\begin{tabular}{lccc}
Material& $|t|$ (eV)  & $d$ (nm) & ${\mu_0t e^2}/{\hbar^2 d}$ ($\mu\mathrm{m}^{-2}$)  \\
\hline
YBCO  & 0.38~\cite{YBCO_TB} &0.585~\cite{YBCOLC}& 302\\
Bi2212 &0.45~\cite{Bi2212_TB}& 0.77~\cite{Bi2212LC}& 271\\
Hg1201& 0.54~\cite{Bi2212_TB}&0.95~\cite{Hg1201LC}& 264\\
LSCO   & 0.48~\cite{LSCO_TB}& 0.66~\cite{LSCOLC}&337 \\
Tl2201 & 0.43~\cite{Pavarini2001}&1.16 ~\cite{Tl2201LC}&172 \\
NaCCOC&0.43~\cite{NaCCOC_TB}& 0.75~\cite{CCOCLC}&266 \\
\end{tabular}
\end{ruledtabular}
\end{table}

Experimental studies sometimes report the $\mu$SR relaxation rate $\sigma$, which is related to the inverse squared penetration depth via~\cite{Bernhard,muSR1,muSR2,muSR3}
\begin{equation}
\sigma[\mu\mathrm{s}^{-1}]
=
7.086\times10^4\,
\lambda^{-2}[\mathrm{nm}^{-2}].
\end{equation}

\subsection{The Relation Between Superfluid Stiffness and Antinodal Quasiparticle Weight}

\begin{figure}[t]
  \centering
  \begin{minipage}[t]{0.47\linewidth}
    \raggedright
    \textbf{(a)}\par
    \noindent\includegraphics[width=\linewidth,trim={7pt 7pt 7pt 7pt},clip]{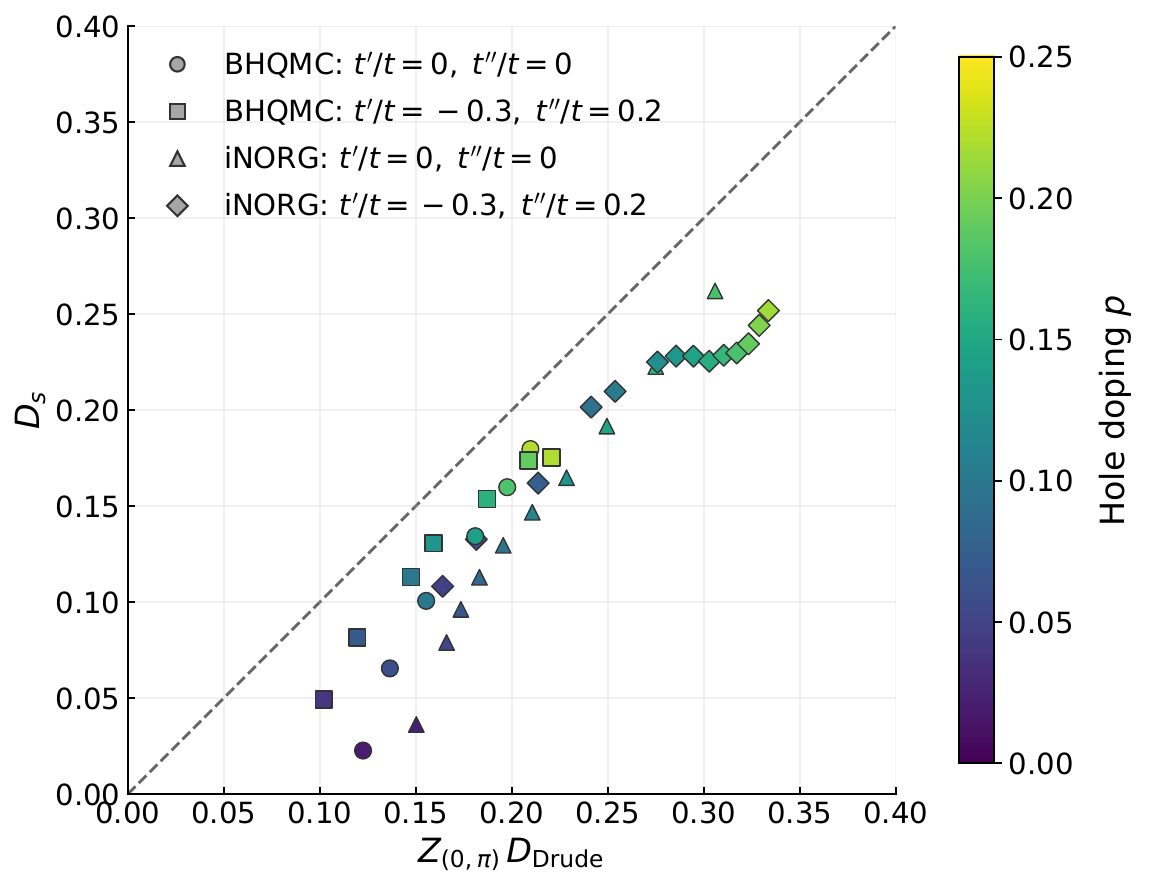}
  \end{minipage}\hfill%
  \begin{minipage}[t]{0.49\linewidth}
    \raggedright
    \textbf{(b)}\par
    \noindent\includegraphics[width=\linewidth]{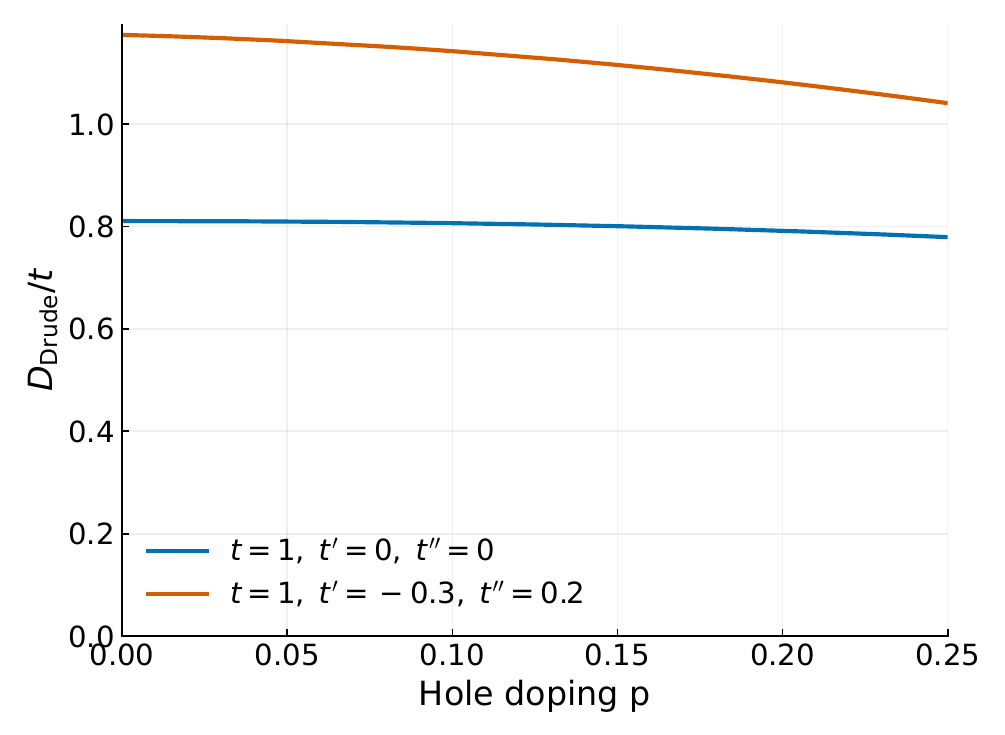}
  \end{minipage}
   \caption{(a) Low-temperature superfluid stiffness versus the product $Z_{(0,\pi)}D_{\rm Drude}$ of the antinodal quasiparticle weight and the band Drude weight for both parameter sets, calculated with the two impurity solvers. BHQMC results are taken at $T=0.01$ and iNORG results at $T=0$. (b) Band Drude weight $D_{\rm Drude}$ of the two tight-binding models as a function of hole doping.}
   \label{stiffness_vs_Z}
 \end{figure}

In an effective-mass picture, $D_s$ scales as $n_s/m^*$. If the
superfluid fraction $n_s/n$ varies weakly with doping, this gives
$D_s\propto D_{\mathrm{Drude}}m_{\mathrm{band}}/m^*$, where
$D_{\mathrm{Drude}}$ is the noninteracting band Drude weight. Using
the antinodal quasiparticle weight $Z_{(0,\pi)}$ as a proxy for the
inverse mass enhancement $m_{\mathrm{band}}/m^*$, we compare the
low-temperature stiffness from Figs.~\ref{T0stiffness} and
\ref{DsT}(a),(b) with $Z_{(0,\pi)}D_{\mathrm{Drude}}$ in
Fig.~\ref{stiffness_vs_Z}(a) for both hopping parameter sets and
impurity solvers.
We calculate the band Drude weight and the antinodal quasiparticle
weight as
\begin{align}
D_{\mathrm{Drude}} &=
    \frac{2}{N}\sum_{\vk}
    \frac{\partial^2\epsilon_{\vk}}{\partial k_x^2}
    \,\Theta(\mu-\epsilon_{\vk}),\\
Z_{(0,\pi)} &=
    \left[1-\left.
    \frac{\partial\operatorname{Im}\Sigma_{(0,\pi)}(i\omega)}
    {\partial\omega}\right|_{\omega=0^+}\right]^{-1},
\end{align}
where $\mu$ is the chemical potential, $\epsilon_{\vk}$ is the
tight-binding dispersion, and $N$ is the number of momentum points.
We extract the zero-frequency slope from a polynomial fit to the
imaginary part of the normal antinodal self-energy at the lowest
positive Matsubara frequencies. The extrapolated intercept is
negligible at the temperatures considered in superconducting state.

We note that in the clean weak-coupling BCS limit at $T=0$, $D_s$ equals the
bare-band Drude weight shown in Fig.~\ref{stiffness_vs_Z}(b) and
is substantially larger than the interacting stiffness in
Fig.~\ref{T0stiffness} of the main text.

Figure~\ref{stiffness_vs_Z}(a) shows an overall positive correlation
between $D_s$ and $Z_{(0,\pi)}D_{\mathrm{Drude}}$. For each hopping
parameter set, the bare-band Drude weight varies only weakly over
$0<p<0.25$ [Fig.~\ref{stiffness_vs_Z}(b)], so most of the doping
dependence of the product comes from $Z_{(0,\pi)}$. The correlation
thus points to quasiparticle mass renormalization as an important
contributor to the suppression of stiffness toward the Mott state.

The largest relative deviations occur at the smallest stiffnesses,
where $D_s$ is more strongly suppressed than the overall trend
suggests. The comparison is qualitative: $Z_{(0,\pi)}$ probes a
single momentum sector, whereas $D_s$ depends on the full
momentum-dependent electronic structure and current vertex.

\subsection{Computational Details}

\paragraph{iNORG impurity solver.}
At zero temperature, the plaquette impurity problem is solved with iNORG~\cite{iNORG}, a Lanczos-based solver in which the natural orbitals renormalization group (NORG)~\cite{NORG} truncates the many-body Hilbert space. The solver has been compared with continuous-time quantum Monte Carlo~\cite{Zen2025} and used in earlier multiorbital calculations~\cite{Tian2024,Chen2024,Wang2025}. The impurity problem is formulated in the cluster-momentum basis of the $2\times2$ plaquette and in Nambu form, which gives eight impurity spin-orbitals. The hole (spin-down) block is particle-hole transformed, so that for singlet pairing the Nambu Hamiltonian conserves the Nambu particle number and the occupancy constraints described below apply directly. The continuous hybridization function, an $8\times8$ Nambu matrix $\Delta_{\rm hyb}(i\omega_n)$ that includes the anomalous blocks, is represented by 24 bath spin-orbitals, six per cluster momentum, so that the NORG problem contains 32 spin-orbitals. The bath energies and the normal and anomalous hybridization amplitudes are independent fit parameters, and no symmetry beyond the cluster-momentum block structure is imposed on them. They are determined by a least-squares fit, with equal weight for all frequencies, of all matrix elements of $\Delta_{\rm hyb}(i\omega_n)$ on a Matsubara grid with an effective inverse temperature $\beta_{\rm eff}=100\pi/t\approx314/t$, using the 375 lowest frequencies ($\omega_n<7.5t$). At $\mathbf A=0$ the four momentum sectors decouple and are fitted independently. At finite $\mathbf A$ the Peierls phase couples the momentum sectors in pairs; each coupled pair is fitted jointly with 12 bath spin-orbitals, so that the total number of bath spin-orbitals stays at 24. The self-consistent loop at finite $\mathbf A$ starts from the converged $\Sigma(\mathbf A=0)$ with a newly fitted bath. 

In the NORG iteration, the bath orbitals are rotated into the natural orbitals of the bath block of the ground-state single-particle density matrix, while the impurity orbitals are kept fixed. Twelve orbitals, the eight impurity orbitals and the four bath natural orbitals with occupations farthest from zero and one, are active with unrestricted occupation. The remaining 20 bath natural orbitals form a nearly full group and a nearly empty group of ten orbitals each, in which the total number of excitations, holes in the nearly full group plus electrons in the nearly empty group, is limited to five. The Nambu particle number is fixed at 16, half the number of spin-orbitals (the $S_z=0$ sector), and the resulting Hilbert space has dimension $1.35\times10^{7}$, compared with $6.0\times10^{8}$ for the unrestricted sector. The ground state is obtained with the Lanczos algorithm, and the orbital rotation and the restricted space are updated until the relative change of the ground-state energy between successive iterations falls below $10^{-10}$. The error bars on the zero-temperature stiffness are estimated from the bath-fitting accuracy. The impurity Green's function is then evaluated with a Krylov-space (Lanczos) method in the $N\pm1$ sectors on the same Matsubara grid, and the self-energy follows from the Dyson equation. The cDMFT loop is converged when the weighted relative change of the self-energy between successive iterations falls below $10^{-4}$.

\paragraph{BHQMC impurity solver.}
The BHQMC solver is based on a dressed strong-coupling hybridization expansion, which we truncate at fourth order in this work. As discussed in our previous work~\cite{BHQMC}, Bulla's trick~\cite{Bulla} evaluates the self-energy by sampling an additional vertex function and produces results that are less noisy than that obtained from the Dyson equation. However, if the expansion order is not high enough for convergence, Bulla's self-energy is not fully consistent with the truncated expansion functional. Therefore, to ensure that the self-consistent DMFT calculation is conserving, the self-energy used in the self-consistency loop must be obtained by the Dyson equation. In post-processing calculations, including those of the order parameter and superfluid stiffness, we instead use the self-energy obtained with Bulla's~\cite{Bulla} or Kugler's~\cite{Kugler} method.

\begin{figure}[t]
  \centering   \includegraphics[width=0.9\columnwidth,trim={7pt 7pt 7pt 7pt},clip]{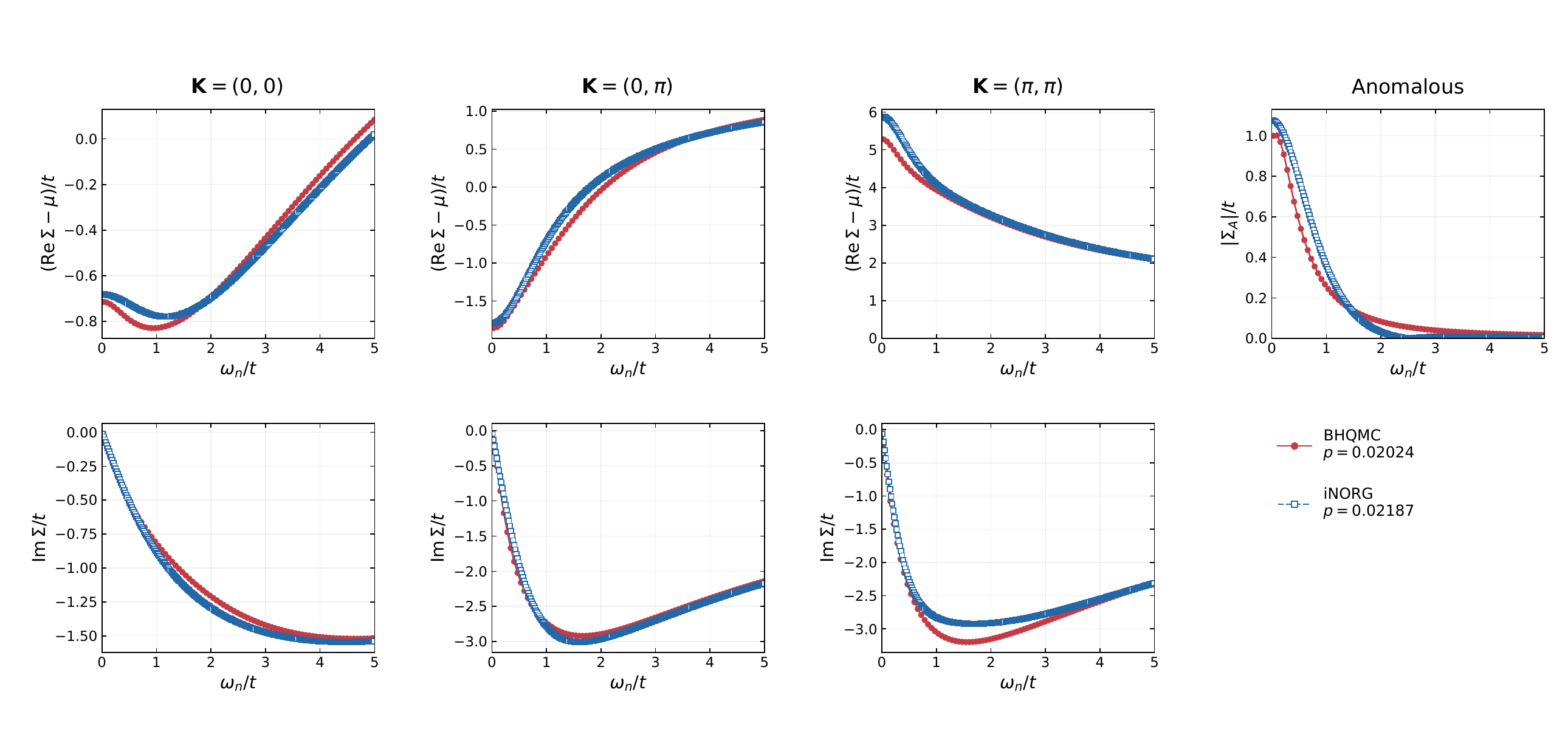}
   \caption{Self-energy from the iNORG impurity solver at $T=0$ and from the BHQMC impurity solver at $\beta=100$ at about $2\%$ hole doping in the nearest-neighbor Hubbard model.}
   \label{sigmabench}
 \end{figure}

Figure~\ref{sigmabench} compares the fourth-order BHQMC self-energy
at $\beta=100$ with the $T=0$ iNORG result for the nearest-neighbor
Hubbard model at about $2\%$ hole doping. The agreement in both the
normal and anomalous components supports the use of the fourth-order
expansion at this doping. We estimate the uncertainty in the
finite-temperature stiffness in Fig.~\ref{DsT} from its sensitivity
to the choice of self-energy estimator. 

\begin{figure}[htbp]
  \centering

  \begin{minipage}[t]{0.49\linewidth}
    \centering
    \llap{\raisebox{-0.2em}{\textbf{(a)}\hspace{-0.5em}}}%
    \includegraphics[
      width=\linewidth,
      trim={6pt 6pt 6pt 6pt},
      clip
    ]{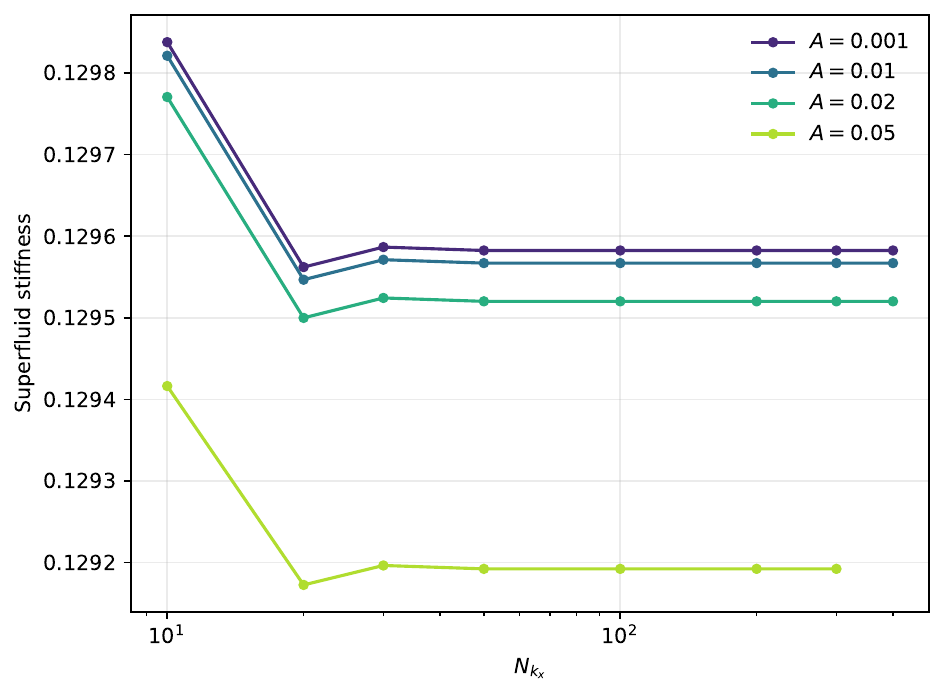}
  \end{minipage}\hfill%
  \begin{minipage}[t]{0.49\linewidth}
    \centering
    \llap{\raisebox{-0.2em}{\textbf{(b)}\hspace{-0.5em}}}%
    \includegraphics[
      width=\linewidth,
      trim={6pt 6pt 6pt 6pt},
      clip
    ]{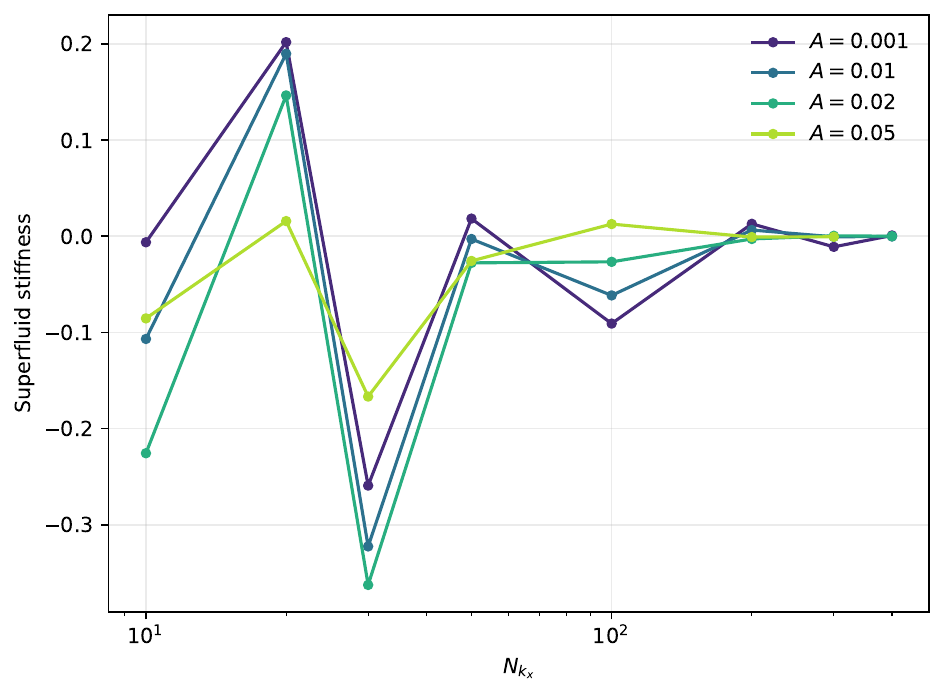}
  \end{minipage}

  \caption{
    Superfluid stiffness of the nearest-neighbor Hubbard model as a function of $x$-direction $k$-point density and value of phase gradient $A$, computed with the iNORG self-energy at (a) $9.8\%$ doping, in the superconducting state, and (b) $23\%$ doping, in the normal state.
  }
  \label{finitesize}
\end{figure}

\paragraph{$k$-mesh convergence of the superfluid stiffness.}
To avoid systematic errors arising from insufficient $k$-point sampling, we examine the convergence of the superfluid stiffness with respect to the momentum-space mesh. When the phase gradient $\vA=A\hat{x}$ is applied, the number of $k$ points along the $x$ direction, $N_{k_x}$, requires more careful convergence, whereas $N_{k_y}=30$ is sufficient. As shown in Fig.~\ref{finitesize}(a), $N_{k_x}=30$ and $A=0.01$ give a relative error of approximately $0.1\%$ for states near the center of the superconducting dome. However, overdoped states that are weakly superconducting or nonsuperconducting require substantially larger values of $N_{k_x}$ since the superconducting gap is mostly or completely closed, particularly when $A$ is small. At the normal-state point shown in Fig.~\ref{finitesize}(b), the systematic error for $N_{k_x}=30$ is of the same order as the largest superfluid stiffness found in this study, and the stiffness approaches zero only at $N_{k_x}=400$ for $A=0.001$. To reduce this error for weakly superconducting states, we use $N_{k_x}=300$ and $A=0.01$ in all superfluid-stiffness calculations, which remains feasible at a modest computational cost. 

\twocolumngrid

\end{document}